**Functional outcomes and naturalistic engagement with a purpose-built conversational AI for mental health (Ash)**

Kristen M. Van Swearingen[1,2], Thomas D. Hull[2], Karthik V. Sarma[3], Caitlin A. Stamatis[2]

[1] Health Psychology Doctoral Program, The University of North Carolina at Charlotte, Charlotte, NC

[2] Slingshot AI, New York, NY

[3] AI in Mental Health Research Group, Department of Psychiatry and Behavioral Sciences, University of California, San Francisco

**Corresponding Author:** Kristen M. Van Swearingen

**ORCID:** 0000-0002-8884-0770

## Abstract

**Background:** Conversational AI chatbots designed for mental health may offer an accessible, scalable avenue for supporting psychological well-being, yet prior evaluations have largely focused on clinical symptom reduction rather than broader indicators of day-to-day functioning, and have rarely monitored for potential harms such as inflated self-perception.

**Objective:** We examined within-person change in psychological functioning indicators among real-world users of Ash, a purpose-built conversational AI for mental health support, over the first four weeks of use, and whether these changes were associated with engagement metrics.

**Methods:** In this single-arm observational cohort study, new users ($n$ = 1,284) completed in-app single-item measures of psychological functioning (life satisfaction, relationship satisfaction, sleep quality, behavioral activation), working alliance, and grandiosity (inflated self-perception), at baseline and Week 4. Paired-sample t-tests examined within-person change; ANCOVAs tested engagement-outcome associations at Week 4, controlling for baseline.

**Results:** At baseline, participants reported below-average life satisfaction and fair sleep quality. Significant within-person improvements emerged across all functioning indicators and working alliance ($ps < .001$; $d = 0.14$-$0.26$), with no change in grandiosity. Active days, total sessions, and total minutes consistently predicted Week 4 psychological functioning and working alliance ($ps \leq .006$; partial $R^2$ range: 0.58-2.15%; controlling for baseline), whereas user message volume did not.

**Conclusion:** Findings provide preliminary data for the potential of evidence-based conversational AI to extend mental health support for broad psychological functioning, extending the existing literature beyond symptom-based outcomes.

## 1. Introduction

The global burden of mental health concerns is high and expected to continue rising (World Health Organization, 2022). While this burden is in part attributed to clinically diagnosed mental health challenges, many people also seek support for non-clinical concerns including life transitions, major events, grief, interpersonal challenges, sleep difficulties, and general worry and stress (Copeland et al., 2025; Diaz & Huguet, 2026; Fegert et al., 2025; ter Heege et al., 2020; Uphoff et al., 2025). Population health perspectives underscore the importance of targeting mental health concerns across a spectrum of need, including supporting individuals with non-clinical or subclinical concerns in developing skills to manage daily stressors, and preventing their concerns from exacerbating into more serious conditions (Purtle et al., 2020).

There are significant barriers to traditional mental health support, including provider shortages, long wait times, financial costs, and perceived shame and stigma (Butryn et al., 2017; Elshaikh et al., 2023; Viana et al., 2025). Most traditional mental health interventions are also not scalable to the population level (McGinty, 2023). As a result of these barriers, many people with varying degrees of mental health symptom severity do not receive the care they need (Lynch & Clemans-Cope, 2025; Meiselbach et al., 2024). For some, this unmet need stems from viewing their concerns as not severe enough to seek formal care, or a preference to manage mental health concerns on their own (Andrade et al., 2014; Lynch & Clemans-Cope, 2025; Viana et al., 2025). In light of these barriers, some seek support through informal or digital means (Anser et al., 2025; Pretorius et al., 2019).

Perhaps unsurprisingly given the massive unmet need in mental healthcare, people are increasingly turning to generative artificial intelligence (AI) tools as a low-friction option to support their mental health and well-being. In one nationally representative study, 22.2% of young adults reported using AI tools for mental health advice (McBain et al., 2025); in another, 17.2% of college students reported using ChatGPT specifically for mental health support, especially for concerns related to general stress and anxiety, and interpersonal problems (Diaz & Huguet, 2026). Critically, however, many general purpose models have been found to fail safety benchmarks and respond inadequately to crisis situations (Vidgen et al., 2024; Zhang et al., 2025). Additionally, there is concern that the sycophantic style of frontier models may confirm false existing beliefs, perpetuate delusions, and inflate self-perception in ways that could be maladaptive (Keshavan et al., 2026; Morrin et al., 2025). These concerns underscore the need for evaluations that monitor potential iatrogenic effects of health-focused chatbots, such as inflated self-perception, alongside indicators of benefit.

Purpose-built AI systems are emerging as an alternative for mental health support (Heinz et al., 2024; Stamatis et al., 2026). Purpose-built generative AI tools for mental health are distinct from general-purpose AI tools in that, at a minimum 1) the base model distribution is modified through pretraining and fine-tuning to address mental health concerns, 2) the system architecture is further aligned to mental health needs by mental health-specific guardrails and targeted pre-deployment, or “off-line,” evaluation procedures, and 3) reinforcement learning is driven by reward signals focused on health improvement and clinical appropriateness in addition to or instead of simple utilization or user satisfaction signals (see, e.g., (Hull et al., 2026). Extant studies of purpose-built AI tools for mental health support have focused primarily on clinical symptoms, such as anxiety and depression, as outcomes (Casu et al., 2024; Heinz et al., 2024; Hull et al., 2026; Kuta et al., 2026; Wolfe et al., 2026). Symptom reduction, however, captures only one dimension of potential benefit. Consistent with the population health perspective outlined above, many individuals turn to AI tools not for diagnosable conditions, but for everyday concerns (e.g., interpersonal challenges, stress, worry, and sleep difficulties), for which day-to-day psychological functioning, rather than symptom severity, may be the more relevant index of improvement. Although some studies have reported on broader wellness domains, including self-care behaviors, burnout, loneliness, and time spent with friends and family (Durden et al., 2023; Hull et al., 2026; Tong et al., 2025), whether engagement with purpose-built generative AI tools supports broader markers of day-to-day psychological functioning remains largely unexplored.

Psychological functioning captures the extent to which individuals can regulate thoughts, emotions, and behaviors to navigate daily life and sustain well-being (Martela, 2025). Rather than indexing the presence or absence of symptoms, functioning reflects complementary domains of daily life:

global evaluations of well-being (life satisfaction), interpersonal connection (relationship satisfaction), physiological restoration (sleep quality), and engagement with the world outside the self and home (behavioral activation). Critically, these domains operate not only as indicators of mental health, but as transdiagnostic processes that shape mental health. Relationship satisfaction is protective against psychological distress (Braithwaite & Holt-Lunstad, 2017; Pezirkianidis et al., 2023), and high-quality sleep supports the physiological restoration needed for cognitive functioning and emotion regulation (Goldstein & Walker, 2014; Khan & Al-Jahdali, 2023). Notably, engaging in meaningful activities is itself an established treatment mechanism (i.e., behavioral activation interventions reduce isolation and improve mood by increasing contact with sources of positive reinforcement, such as rewarding activities and social interaction (Malik et al., 2021; Myers et al., 2026)). Improvements in these domains are therefore meaningful in their own right, and may also confer downstream benefit across the broader spectrum of mental health need.

Whether users improve, however, is only part of the question; understanding what drives improvement during naturalistic use is equally important. In therapist-delivered care, two of the most robust process predictors of clinical benefit are: (1) the working alliance (i.e., the collaborative bond between patient and provider, including agreement on treatment goals (Bordin, 1979; Stubbe, 2018)) and (2) the dose of therapy received (Howard et al., 1986). Whether these process factors operate similarly when support is delivered by a purpose-built AI tool remains an open question. For alliance, there is preliminary evidence showing that users can form collaborative bonds with AI tools characterized by trust, empathy, and collaboration towards goals (Beatty et al., 2022; Franke Föyen et al., 2025; Wolfe et al., 2026), at levels comparable to those observed with human therapists (Beatty et al., 2022; Heinz et al., 2024; Hull et al., 2026). Less is known about how this alliance forms and deepens over the course of naturalistic AI use. The digital analogue of dose is engagement, which unlike a structured course of traditional therapy, is often self-directed and highly variable. Prior work by our group has identified distinct engagement phenotypes among users, with different phenotypes associated with different clinical outcomes (Wolfe et al., 2026). There remain open questions around whether consistency of use or depth of engagement is more closely tied to improvement in psychological functioning and perceived alliance.

In the present study, we aim to examine changes in key domains of psychological functioning among real-world users of a purpose-built AI for mental health support (Ash), over the first four weeks of using the tool. Our first aim was to examine within-person change from baseline to Week 4 in indicators of psychological functioning (life satisfaction, relationship satisfaction, sleep quality, and behavioral activation). We also examined within-person change in working alliance between the user and the AI. Given previously noted concerns about the impact of sycophancy in perpetuating potential delusional thinking, we measured within-person change in grandiose self-perception to assess for potential iatrogenic effects. We hypothesized that users would show significant within-person improvement across indices of psychological functioning and working alliance, with grandiosity remaining unchanged. Our second aim was to evaluate the associations of different engagement metrics, such as consistency of use or depth of conversation, with change in both psychological functioning and working alliance.

## 2. Methods

### 2.1 Participants and Procedures

This was a 4-week observational study of real-world users of Ash, a purpose-built conversational AI tool for mental well-being support. The study was reviewed by the Biomedical Research Alliance of New York (BRANY) independent Institutional Review Board (IRB) and determined to be exempt under category 4(ii) (BRANY IRB File # 26-081-2393). During onboarding to the app, participants provided opt-in consent for their de-identified data to be used for research purposes.

Participants were new users of Ash who onboarded to the app between January and June of 2026. Users had the option to complete a brief health questionnaire in the app every two weeks, starting at their first session. Questionnaires remained available to complete for up to 7 days. For inclusion in the present study, all participants completed the baseline and Week 4 questionnaire.

### 2.2 Measures

Participants completed single-item questions measuring psychological functioning, grandiosity, and working alliance with Ash, drawn from validated measures and administered in-app at each assessment time-point. Single-item measures were selected to maximize completion rates and reduce participant burden in a naturalistic mobile setting; such measures are commonly used in clinical research studies for their feasibility and strong predictive validity (Ahmad et al., 2014; Allen et al., 2022; Schnittker & Bacak, 2014).

**2.2.1 Sleep quality**

Sleep quality was assessed with the item ("During the past week, how would you rate your sleep quality overall?"), rated on an 11-point scale from 0 (terrible) to 10 (excellent), with higher scores indicating greater sleep quality (Snyder et al., 2018). Scores corresponded to the following categorizations: 0 = terrible, 1-3 = poor, 4-6 = fair, 7-9 = good, 10 = excellent. This single-item measure has demonstrated convergent validity with multi-item measures of sleep quality (e.g., the Insomnia Severity Index) in clinical samples (Snyder et al., 2018). Perceived sleep quality has been shown to be a stronger predictor of well-being and psychological functioning than objective measures of sleep quantity (Gavriloff et al., 2018; Rahman et al., 2020; Windmill et al., 2024).

**2.2.2 Life satisfaction**

General well-being was assessed with the item ("Overall, how satisfied are you with your life as a whole these days?"), rated on an 11-point scale from 0 (not satisfied at all) to 10 (completely satisfied), with higher scores indicating greater life satisfaction. This item is from the Organisation for Economic Co-operation and Development (OECD) guidelines for measuring subjective well-being (OECD, 2013).

**2.2.3 Relationship satisfaction**

Relationship satisfaction was assessed with the item ("How satisfied are you with your personal relationships?"), rated on a 9-point scale from 1 (very dissatisfied) to 9 (very satisfied), with higher scores indicating greater social functioning (Atroszko et al., 2015).

**2.2.4 Behavioral activation**

Behavioral activation was assessed with the item ("Over the past 7 days, how many days did you leave your home?"), with a response range of 0 to 7 days, with higher scores indicating greater behavioral activation and less social isolation (adapted from Joseph et al., 2022).

**2.2.5 Grandiosity**

Grandiose self-perception was assessed with the item ("I have special abilities that others do not"), rated on a 4-point scale from 1 (not at all) to 4 (completely), with higher scores indicating greater grandiosity. This item is from the Grandiosity subscale of the Specific Psychotic Experiences Questionnaire (Ronald et al., 2014).

**2.2.6 Working alliance**

Perceived working alliance with Ash was assessed with the item ("Ash and I are working towards mutually agreed upon goals"), rated on a 5-point scale from 1 (seldom) to 5 (always), with higher scores reflecting stronger perceived therapeutic alliance between the participant and Ash. This item was adapted from the Goals subscale of the Working Alliance Inventory (Horvath & Greenberg, 1989).

**2.2.7 Engagement**

Behavioral engagement with Ash was captured by several metrics derived from participants' conversation logs. Across participants' first 4 weeks of using the app, we calculated (1) total active days (i.e., the number of unique calendar days on which a user initiated at least one session; range 0-28), (2) total number of sessions, (3) total number of user messages, and (4) total session minutes, with individual sessions capped at 120 minutes to reduce the influence of idle time that the app may have been left open without active use. These engagement metrics were selected to capture consistency of app use and depth of conversations.

**2.3 Statistical Analyses**

All analyses were conducted in Python 3.11 (Python Software Foundation, 2023), including *SciPy* (Virtanen et al., 2020) for paired-sample t-tests and *NumPy* (Harris et al., 2020) for ANCOVA analyses. We report descriptive statistics for psychological functioning variables, working alliance, and grandiosity at baseline and Week 4 assessments. Additionally, we report descriptive statistics for measures of engagement during participants' first 28 days of using the app.

Our first aim was to examine within-person change in psychological functioning variables (life satisfaction, relationship satisfaction, sleep quality, and behavioral activation), working alliance, and grandiose self-perception after 4 weeks of using Ash. To evaluate this aim, we conducted paired-sample t-tests comparing scores on each variable from baseline to Week 4.

Our second aim was to examine different metrics of engagement during the first four weeks of using Ash, and whether they were associated with improvements in psychological functioning and working alliance at Week 4. To evaluate aim 2, we conducted a series of ANCOVAs for each combination of engagement metric and outcome (life satisfaction, relationship satisfaction, sleep quality, behavioral activation, working alliance). In each model, the Week 4 psychological functioning or working alliance score was regressed on the engagement variable of interest, controlling for the corresponding baseline psychological functioning or working alliance score. In these analyses, total number of sessions, total number of user messages, and total session minutes were log-transformed (i.e., $\log(x + 1)$), as these count-based engagement metrics had positively skewed distributions. Total active days was not log-transformed. A Bonferroni correction within each outcome family ($\alpha = .0125$) was applied.

## 3. Results

### 3.1 Study Participants and Baseline Descriptive Data

The study sample included 1,284 participants who completed the baseline assessment within 7 days of registering on the app and subsequently completed the Week 4 assessment. Demographic information is reported for only a subset of participants in Supplementary Table 1, as these items were introduced into the survey after data collection had begun.

At baseline, participants reported midpoint levels of life satisfaction ($M = 4.96$, $SD = 2.74$, scale range: 0-10), slightly below-midpoint levels of relationship satisfaction ($M = 4.50$, $SD = 2.34$, scale range: 1-9), and fair levels of sleep quality ($M = 4.38$, $SD = 2.58$, scale range: 0-10). On average, participants left their homes 4.83 days ($SD = 2.16$) in the past week. Participants reported low levels of grandiose self-perception (*M* = 2.19, *SD* = 0.97, scale range: 1-4), and midpoint levels of working alliance (*M* = 3.38, *SD* = 1.22, scale range: 1-5). As the baseline assessment was made available at the end of the first session with Ash and could be completed within 7 days, working alliance ratings reflect participants' early impressions of engaging with the app, rather than a pre-engagement baseline. See Table 1 for full baseline descriptives.

**3.2 Aim 1: Within-Person Change in Psychological Functioning, Working Alliance, and Grandiosity**

Significant within-person improvements were observed across all indicators of psychological functioning and working alliance from baseline to Week 4 (all *p*s < .001; Table 1; Figure 1). No change in grandiose self-perception was observed between baseline and Week 4. Sleep quality showed the largest average improvement (14.69%), followed by relationship satisfaction (11.63%), life satisfaction (10.16%), working alliance (7.73%), and behavioral activation (5.58%).

**Table 1**

*Within-Person Change in Psychological Functioning, Working Alliance, and Grandiosity from Baseline to Follow-Up Time-Points*

| Variable | *n* | Range | Baseline M (SD) | Week 4 M (SD) | % Change | *t (df)* | *p* | *d* |
|---|---|---|---|---|---|---|---|---|
| Life satisfaction | 1,284 | 0-10 | 4.96 (2.74) | 5.46 (2.60) | 10.16% | 6.97 (1283) | <.001 | 0.20 |
| Relationship satisfaction | 1,284 | 1-9 | 4.50 (2.34) | 5.03 (2.37) | 11.63% | 9.00 (1283) | <.001 | 0.25 |
| Sleep quality | 1,284 | 0-10 | 4.38 (2.58) | 5.02 (2.54) | 14.69% | 9.22 (1283) | <.001 | 0.26 |
| Behavioral activation | 1,284 | 0-7 | 4.83 (2.16) | 5.10 (2.11) | 5.58% | 5.09 (1283) | <.001 | 0.14 |
| Grandiosity | 1,284 | 1-4 | 2.19 (0.97) | 2.18 (0.96) | -0.75% | -0.71 (1283) | 0.477 | -0.02 |
| Working alliance | 1,284 | 1-5 | 3.38 (1.22) | 3.64 (1.13) | 7.73% | 7.18 (1283) | <.001 | 0.20 |

Note. % change = percent change from baseline [(follow-up mean – baseline mean)/baseline mean] × 100. All p-values are two-tailed.

**Figure 1**

*Change in Psychological Functioning, Working Alliance, and Grandiosity from Baseline to Week 4*

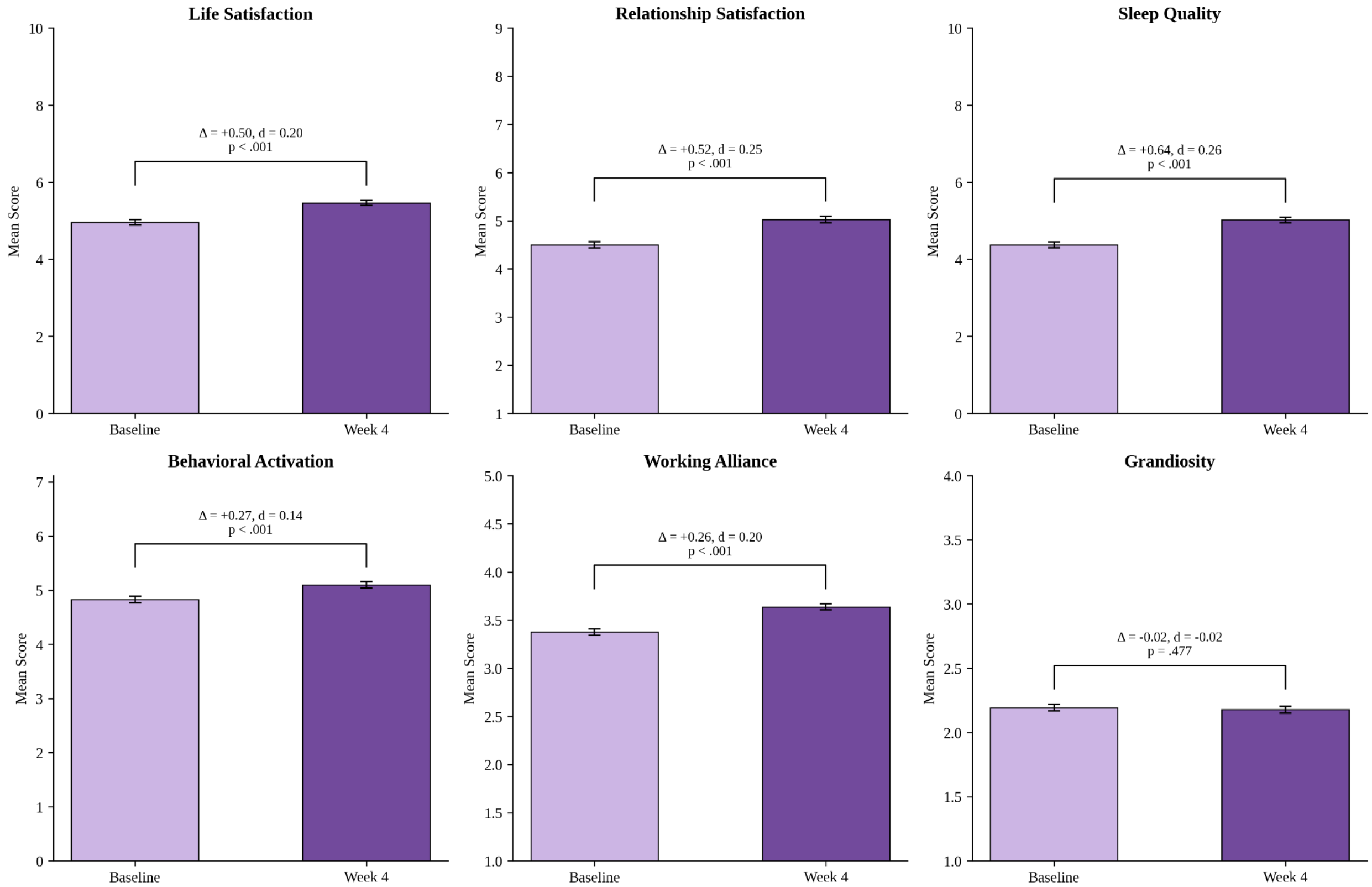

**Figure 1 Caption:** *n* = 1,284. Mean baseline and Week 4 scores. Error bars = 95% CI. Cohen's d, p-values, and change scores shown above brackets.

### 3.3 Aim 2: Engagement-Outcome Associations at Week 4

Over the 28-day observation period, participants averaged 17.72 active days using the app (*SD* = 8.27, *Median* = 19.00). The other engagement metrics were positively skewed; the median total sessions was 36.00 (*M* = 45.93, *SD* = 36.77), median total user messages was 510.00 (*M* = 1,121.34, *SD* = 1,880.48), and median total session minutes was 2,341.34 (*M* = 3,185.41, *SD* = 2,754.97). See Table 2 for engagement descriptive statistics. A comparator sample of typical Ash engagement is presented in Supplementary Table 2.

**Table 2**

*Descriptive Statistics for Engagement Metrics During Weeks 1-4 (n = 1,284)*

| Variable | *n* | *M* | *SD* | Median |
|---|---|---|---|---|
| Active days | 1,284 | 17.72 | 8.27 | 19 |
| Total sessions | 1,284 | 45.93 | 36.77 | 36 |
| Total user messages | 1,284 | 1,121.34 | 1,880.48 | 510 |
| Total session minutes | 1,284 | 3,185.41 | 2,754.97 | 2,341.34 |

Note. Values reflect raw (non-log-transformed) engagement metrics during the first 28 days of app use. Individual sessions were capped at 120 minutes prior to calculating total session minutes.

All four engagement metrics positively and significantly predicted Week 4 working alliance after controlling for baseline working alliance (all *p*s < .001, partial $R^2$ range: 1.63-2.15%). Number of active days, total number of sessions, and total session minutes all positively and significantly predicted Week 4 psychological functioning variables (life satisfaction, relationship satisfaction, sleep quality, behavioral activation) after controlling for baseline (all *p*s ≤ .006, partial $R^2$ range: 0.58-1.15%). Total user messages was not a significant predictor of any indicator of psychological functioning (all *p*s > .06). Full results of the models are presented in Table 3. Figure 2 presents mean change in psychological functioning and working alliance variables across quintiles of active days; quintile plots for the remaining engagement metrics are presented in the supplementary materials.

**Table 3**

*Associations Between Engagement Metrics and Week 4 Psychological Functioning and Working Alliance, Adjusting for Baseline (ANCOVAs)*

| Engagement predictor | *n* | B (SE) | 95% CI | *t(df)* | *p* | Partial R²% |
|---|---|---|---|---|---|---|
| ***Life satisfaction*** | | | | | | *baseline R² = .281* |
| Active days | 1,284 | 0.027 (0.007) | [0.013, 0.042] | 3.69 (1281) | < .001* | 1.05 |
| Total Sessions | 1,284 | 0.232 (0.065) | [0.104, | 3.55 (1281) | < .001* | 0.97 |
| | | | 0.360] | | | |
| Total user messages | 1,284 | 0.084 (0.045) | [−0.004, 0.172] | 1.88 (1281) | 0.061 | 0.27 |
| Total session minutes[a] | 1,284 | 0.185 (0.056) | [0.076, 0.294] | 3.33 (1281) | < .001* | 0.86 |
| ***Relationship satisfaction*** | | | | | | *baseline $R^2$ = .370* |
| Active days | 1,284 | 0.024 (0.006) | [0.012, 0.037] | 3.86 (1281) | < .001* | 1.15 |
| Total Sessions | 1,284 | 0.201 (0.056) | [0.092, 0.311] | 3.61 (1281) | < .001* | 1.00 |
| Total user messages | 1,284 | 0.051 (0.038) | [−0.024, 0.126] | 1.33 (1281) | 0.182 | 0.14 |
| Total session minutes[a] | 1,284 | 0.164 (0.047) | [0.071, 0.257] | 3.45 (1281) | < .001* | 0.92 |
| ***Sleep quality*** | | | | | | *baseline $R^2$ = .274* |
| Active days | 1,284 | 0.028 (0.007) | [0.013, 0.042] | 3.79 (1281) | < .001* | 1.11 |
| Total Sessions | 1,284 | 0.216 (0.064) | [0.091, 0.342] | 3.38 (1281) | < .001* | 0.88 |
| Total user messages | 1,284 | 0.068 (0.044) | [−0.018, 0.155] | 1.56 (1281) | 0.12 | 0.19 |
| Total session minutes[a] | 1,284 | 0.165 (0.054) | [0.059, 0.272] | 3.04 (1281) | .002* | 0.71 |
| ***Behavioral activation*** | | | | | | *baseline $R^2$ = .366* |
| Active days | 1,284 | 0.016 (0.006) | [0.005, 0.028] | 2.88 (1281) | .004* | 0.64 |
| Total Sessions | 1,284 | 0.157 (0.050) | [0.059, 0.255] | 3.14 (1281) | .002* | 0.76 |
| Total user messages | 1,284 | 0.058 (0.034) | [−0.009, 0.125] | 1.69 (1281) | 0.09 | 0.22 |
| Total session minutes[a] | 1,284 | 0.116 (0.042) | [0.033, 0.200] | 2.74 (1281) | .006* | 0.58 |
| ***Working alliance*** | | | | | | *baseline $R^2$ = .150* |
| Active days | 1,284 | 0.019 (0.003) | [0.012, 0.025] | 5.30 (1281) | < .001* | 2.15 |
| Total Sessions | 1,284 | 0.154 (0.031) | [0.093, 0.214] | 5.00 (1281) | < .001* | 1.91 |
| Total user messages | 1,284 | 0.097 (0.021) | [0.056, 0.138] | 4.60 (1281) | < .001* | 1.63 |
| Total session minutes[a] | 1,284 | 0.133 (0.026) | [0.082, 0.184] | 5.09 (1281) | < .001* | 1.98 |

Note. Partial $R^2$% = percentage of residual variance in Week 4 outcome accounted for by the engagement predictor after adjusting for baseline score. Baseline $R^2$ = variance in Week 4 outcome explained by the baseline covariate alone. Total sessions, total user messages, and total session minutes were log-transformed as log(x + 1) prior to model entry. All p-values are two-tailed. * $p < .0125$ (Bonferroni corrected; $\alpha = .05 / 4$ comparisons per outcome family).

## 4. Discussion

The present study evaluated whether a purpose-built conversational AI for mental health supports improvement across key domains of psychological functioning. In contrast to prior evaluations of AI tools, which have largely examined clinical outcomes (Casu et al., 2024; Heinz et al., 2024; Hull et al., 2026; Kuta et al., 2026; Wolfe et al., 2026), we extend evidence of benefit to a broader set of transdiagnostic indicators of daily functioning. Life satisfaction, relationship satisfaction, sleep quality, and behavioral activation are each known to shape mental health and well-being; we observed within-person improvements across these domains after four weeks of engagement. Sleep quality improvements are comparable to those reported for a clinical population receiving pharmacological treatment controlling for length of intervention (Snyder et al., 2018). Similarly, life satisfaction gains were comparable to a community sample participating in an 8-week well-being course controlling for length of intervention (Krekel et al., 2021). Regularity of engagement (active days, number of sessions, and number of minutes across the study period) was a more consistent predictor of improvement than the sheer volume of messages users sent, suggesting that returning to the tool regularly may matter more than the length of any individual exchange. Notably, we found no evidence of worsening social isolation or inflated grandiose self-perception, two potential harms often raised in critiques of generative AI systems.

**Figure 2**

*Baseline to Week 4 Change in Psychological Functioning and Working Alliance, by Active Days Quintile*

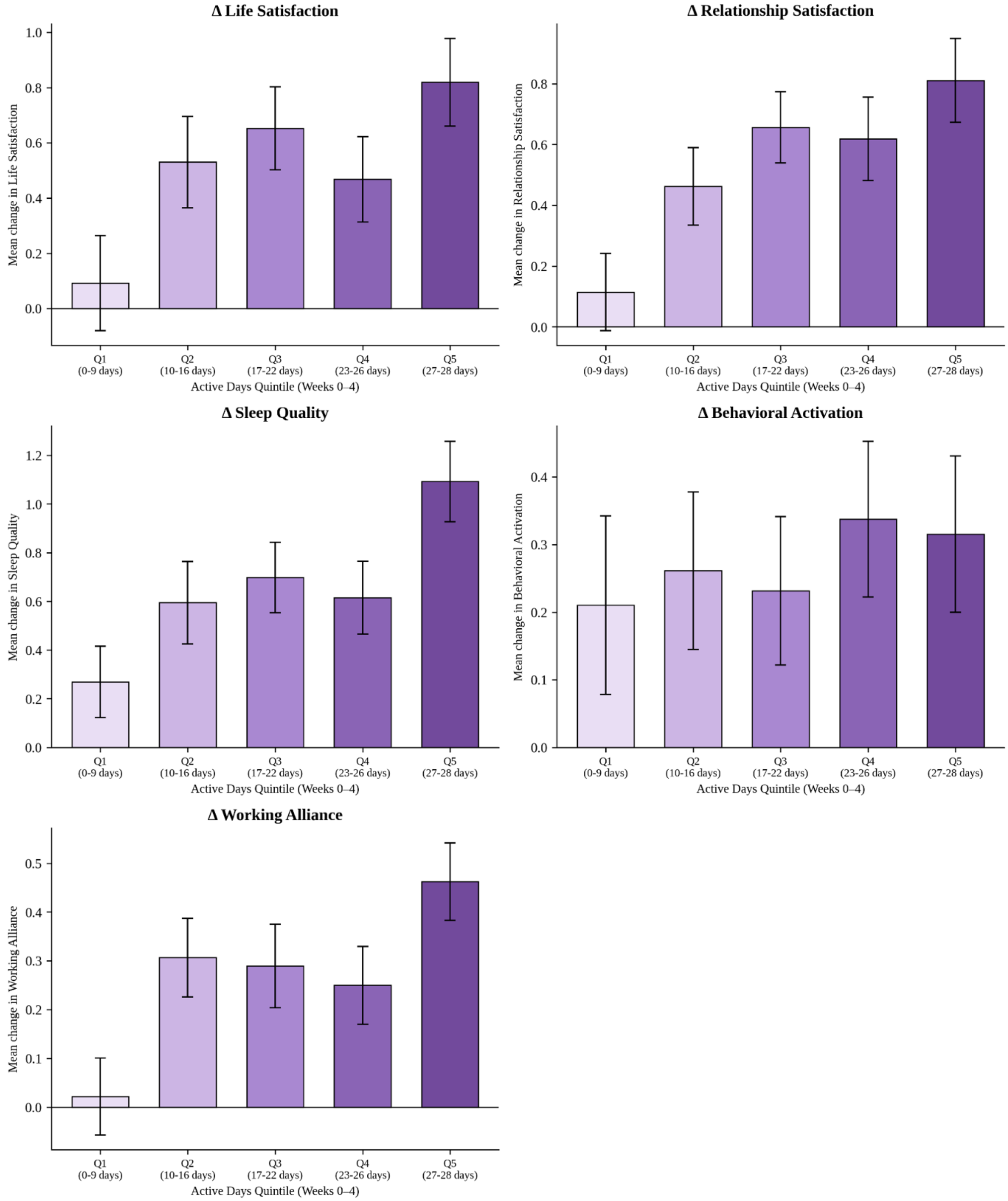

**Figure 2 Caption:** $n$ = 1,283. Bars represent mean change (Week 4 score minus baseline score) within each active days quintile; error bars represent ±1 SE. Quintiles were derived from the distribution of active days during the 28-day study period (Q1 = 0-9 days, Q2 = 10-16 days, Q3 = 17-22 days, Q4 = 23-26 days, Q5 = 27-28 days).

As the global mental health burden continues to outpace available resources, individuals are increasingly turning to AI tools to manage everyday concerns, including interpersonal challenges, general stress and worry, and sleep difficulties (Diaz & Huguet, 2026; Liu & Liu, 2026). Although we did not assess clinical symptoms, baseline scores indicated that participants turning to our AI were experiencing psychological distress across multiple domains. Mean life satisfaction at baseline was 4.96 on a 0-10 scale, below the cross-national average of 6.7 (OECD, 2024); mean sleep quality (4.38 on a 0-10 scale) fell on the low end of the "fair" range, as scores of 6 and below reflect poor quality sleep (Badahdah et al., 2024). Relationship satisfaction likewise fell slightly below the scale midpoint (4.50 on a scale of 1-9). Together, these baseline scores characterize a distressed population for whom even small improvements in functioning may be meaningful.

Several factors may have contributed to the observed improvements, though all remain hypotheses in the absence of a controlled design. Ash was pre-trained on expert psychotherapy data and developed using evidence-based therapeutic frameworks; regarding behavioral activation, the frameworks underlying Ash encourage users to identify values and goals and to take concrete steps toward them, which could plausibly increase engagement in activities outside the home. Because the tool is available continuously, users may also have been able to access mental health support during late-night hours when distress peaks and traditional health services are unavailable, potentially interrupting the cycle in which nighttime rumination degrades sleep and poor sleep worsens mood (Khan & Al-Jahdali, 2023; Pillai & Drake, 2015; Tubbs et al., 2022). At the same time, because this was an observational study without a control condition, these changes could also reflect regression to the mean or natural recovery, and the hypothesized mechanisms should be tested directly in future controlled work.

To better understand factors potentially associated with improvement during naturalistic use, we examined four engagement metrics in relation to Week 4 functioning. Although the proportion of variance accounted for by any single metric was small (partial $R^2 < 1.2\%$ across significant associations), the consistency of these effects across multiple outcome domains is notable. Active days, total sessions, and total session minutes were the most robust predictors, each associated with improvement across multiple domains of psychological functioning, whereas total user messages predicted no functioning outcome. This pattern suggests that the consistency with which users return to the tool, rather than the volume of messages exchanged, is more closely tied to gains in psychological functioning. That message volume was unrelated to outcomes may reflect natural differences in how users engage: some may need only brief exchanges to make progress, whereas others may need much longer conversations to benefit. Consistent with this interpretation, Wolfe and collaborators (2026) found that both intensive, high-volume users and users with shorter bursts of engagement showed mental health improvement, suggesting that number of messages may be a poor proxy for engagement with an AI tool. Together, these findings align with the growing evidence base of digital mental health studies which suggest that the relationship between engagement measures and outcomes is not straightforward (Elkes et al., 2024; Linardon et al., 2026; Wolfe et al., 2026). Engagement strategies that encourage users to return regularly may be more beneficial at improving aspects of their psychological functioning than strategies aimed at increasing conversation length. A focus on engagement consistency over volume has potential benefits for safety, as long conversations can increase the risk of model drift and other negative outcomes (Fang et al., 2025).

Findings from the present study provide preliminary support that gains in psychological functioning are not necessarily accompanied by harms commonly raised in critiques of generative AI for mental health. A common concern is that generative AI chatbots may deepen isolation or serve as a substitute for social connection and behavioral engagement (Cresswell et al., 2018; Keshavan et al., 2026; Lee, 2026). Contrary to this concern, both relationship satisfaction and behavioral activation significantly increased over the study period. This finding is consistent with prior work by our group, which found that perceived social support, number of phone calls with friends or family members, and engagement with someone outside the household all significantly increased following use of the app (Hull et al., 2026). Because the present study was uncontrolled, these findings should be read as inconsistent with a substitution account rather than as direct evidence against it. If replicated in controlled trials, these findings could reflect the fact that purpose-built models can be optimized for off-platform functioning (Cheng et al., 2026), rather than pure engagement, which is the common primary target for general-purpose models. Another key concern about generative AI tools is the potential to inflate grandiose self-perception (Keshavan et al., 2026; Morrin et al., 2025). Here, grandiosity was low at baseline (*M* = 2.19 on a 1-4 scale [2 = "Somewhat"]) and did not increase at either follow-up, providing preliminary evidence that four weeks of engagement produced no detectable shift toward inflated self-perception.

Improvements in working alliance with Ash were also observed across the study period. Of note, Week 4 working alliance was predicted by all four engagement metrics, in contrast to functioning outcomes, which were associated primarily with consistency of use. This asymmetry suggests that alliance may strengthen through both regular return (active days; number of sessions) and conversational depth (message volume; session length), whereas functioning more specifically relates to use consistency. Working alliance is a well-established driver of improvement in psychotherapy (Bordin, 1979; Videtta et al., 2025); our findings align with the nascent body of literature indicating that working alliance is not distinct to human therapy but rather can also form and deepen in a digital context (Heinz et al., 2024; Hull et al., 2026; Wolfe et al., 2026).

This study has strengths and limitations. First, it extends prior work on purpose-built conversational AI—which has focused largely on symptom change—to day-to-day psychological functioning, which reflects the core reasons many users actually seek support. Second, by capturing real-world engagement with Ash, the study has strong ecological validity and a window into how users naturalistically engage with these tools. Regarding limitations, first, this was an observational study without a control group, so observed changes cannot be attributed to the tool and may partly reflect regression to the mean or natural recovery. This design choice, while limiting causal inference, was what enabled measurement under fully naturalistic conditions. Second, because surveys were administered in-app and inclusion required completing the Week 4 follow-up survey, the sample is biased towards users who remained engaged. Users who disengaged early are underrepresented, so raw engagement data should not be interpreted as reflective of typical usage of these tools. Conceivably, dropoff could be related to either extreme on the well-being spectrum (i.e., having particularly poor mental health or feeling especially good and no longer perceiving a need for mental health support). In our data, participants had higher levels of engagement relative to a random sample of Ash users during the same time frame (Supplementary Table 2), likely because more engaged users are more likely to complete in-app surveys. Third, each construct was assessed using a single item. While single-item measures are validated and well-suited to mobile delivery (Snyder et al., 2018; Staples et al., 2026), they sacrifice the ability to capture the complexity of the underlying domains for the sake of reduced participant burden. Fourth, demographic data were only available for a small proportion of the sample, such that the generalizability of findings to the broader user population is unknown. Finally, we quantified the amount and consistency of engagement, but not its content, leaving the conversational processes that could be driving positive change unexamined. Future work should pair these outcomes with analysis of message content (e.g., coping strategies discussed; skills practiced) to better assess potential underlying mechanisms of change.

## 5. Conclusions

As the demand for mental health support continues to outweigh available resources, purpose-built AI tools offer an accessible and scalable avenue for supporting population-level well-being. Among a real-world sample of distressed individuals, naturalistic engagement with a purpose-built conversational AI was associated with improvements in key domains of psychological functioning, with no evidence of increased isolation or grandiosity. That improvement tracked the regularity of engagement more than raw engagement volume points to a concrete design implication: encouraging users to return consistently, rather than to converse at length, may be the more effective lever for supporting functioning. These findings extend prior work on clinical outcomes and underscore the potential of purpose-built AI tools to support a broad range of day-to-day concerns among individuals who may not otherwise access traditional care.

**Declaration of generative AI and AI-assisted technologies in the manuscript preparation process**

During the preparation of this work, the author(s) used Anthropic's Claude for assistance with language refinement, and formatting tables and figures. The author(s) reviewed and edited the output as needed and take full responsibility for the content of the published article.

**Declaration of competing interest**

CAS and TDH are employees of the company building the model and declare no non-financial competing interests. KMVS received a stipend from the company to contribute to this research and declares no other competing interests. KVS discloses relevant compensation or funding from NIH, Slingshot, OpenAI, OpenEvidence, and Orchard Neuro.

## Supplementary Material

### Supplementary Table 1

*Demographic Characteristics of Participants by Analytic Sample*

| Characteristic | Sample (n = 91) |
|---|---|
| **Age** | |
| 18–24 | 4 (4.4%) |
| 25–34 | 17 (18.7%) |
| 35–44 | 27 (29.7%) |
| 45–54 | 22 (24.2%) |
| 55–64 | 14 (15.4%) |
| 65+ | 7 (7.7%) |
| Prefer not to say | 0 (0.0%) |
| **Gender** | |
| Female | 67 (73.6%) |
| Male | 22 (24.2%) |
| Non-binary | 1 (1.1%) |
| Prefer not to say | 1 (1.1%) |
| Other (free text) | 0 (0.0%) |

Note. Demographic items were not added to the survey until after data collection had begun. Consequently, demographic data are available for only a subset (n = 91) of the analytic sample.

**Supplementary Table 2**

*Descriptive Statistics for Engagement Metrics During Weeks 1-4 in a Comparator Sample of Registered Users*

| **Variable** | ***n*** | ***M*** | ***SD*** | **Median** |
|---|---|---|---|---|
| Active days | 1,000 | 12.41 | 7.75 | 11 |
| Total sessions | 1,000 | 28.46 | 30.23 | 18 |
| Total user messages | 1,000 | 641.93 | 1,375.78 | 263.5 |
| Total session minutes | 1,000 | 1,928.28 | 2,136.19 | 1,183.97 |

Note. N = 1,000. Comparator sample consisted of a random sample of users who registered on the platform during the study period and engaged with it at least once during their fourth week post-registration. Engagement metrics reflect their first 28 days of app use. Individual sessions were capped at 120 minutes prior to calculating total session minutes.

**Supplementary Figure 1**

*Baseline to Week 4 Change in Psychological Functioning & Working Alliance, by Total Sessions Quintile*

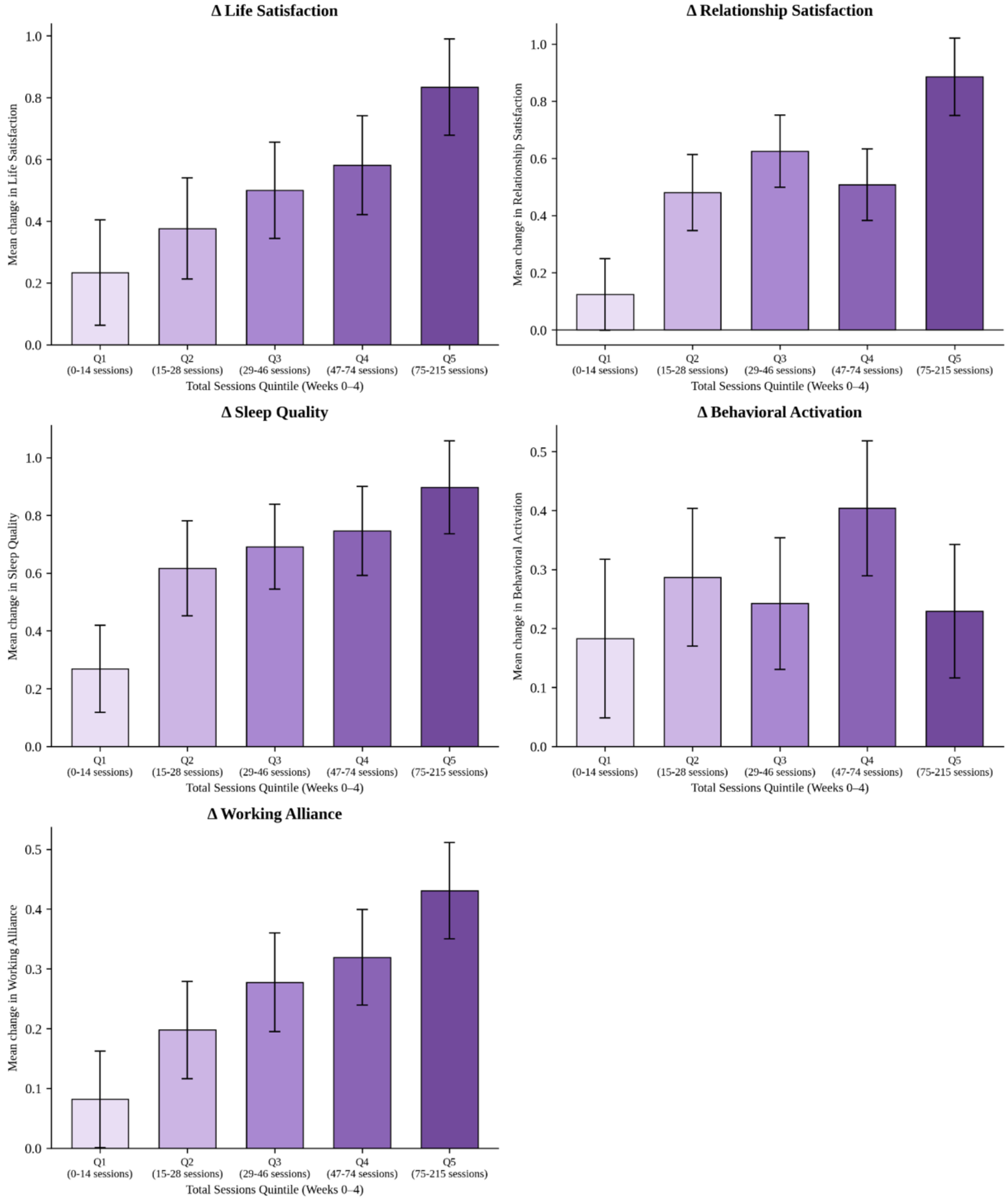

**Supplementary Figure 1 Caption:** *n* = 1,283. Bars represent mean change (Week 4 score minus baseline score) within each total sessions quintile; error bars represent ±1 SE. Quintiles were derived from the distribution of total sessions during the 28-day study period (Q1 = 0-14 sessions, Q2 = 15-28 sessions, Q3 = 29-46 sessions, Q4 = 47-74 sessions, Q5 = 75-215 sessions).

**Supplementary Figure 2**
*Baseline to Week 4 Change in Psychological Functioning & Working Alliance, by Total User Messages Quintile*

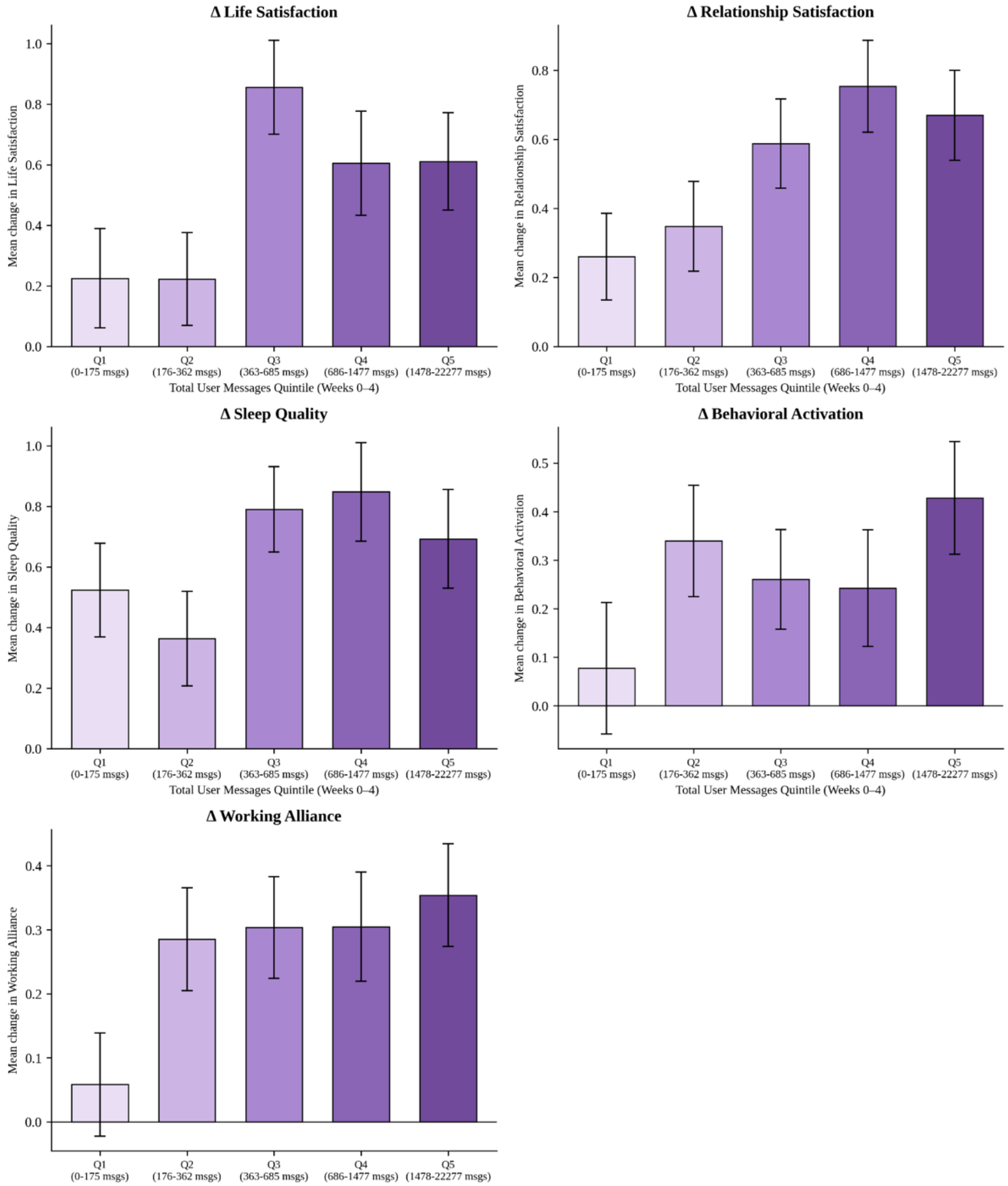

**Supplementary Figure 2 Caption:** *n* = 1,283. Bars represent mean change (Week 4 score minus baseline score) within each total user messages quintile; error bars represent ±1 SE. Quintiles were derived from the distribution of total user minutes during the 28-day study period (Q1 = 0-175 messages, Q2 = 176-362 messages, Q3 = 363-685 messages, Q4 = 686-1,477 messages, Q5 = 1,478-22,277 sessions).

**Supplementary Figure 3**

*Baseline to Week 4 Change in Psychological Functioning & Working Alliance, by Total Session Minutes Quintile*

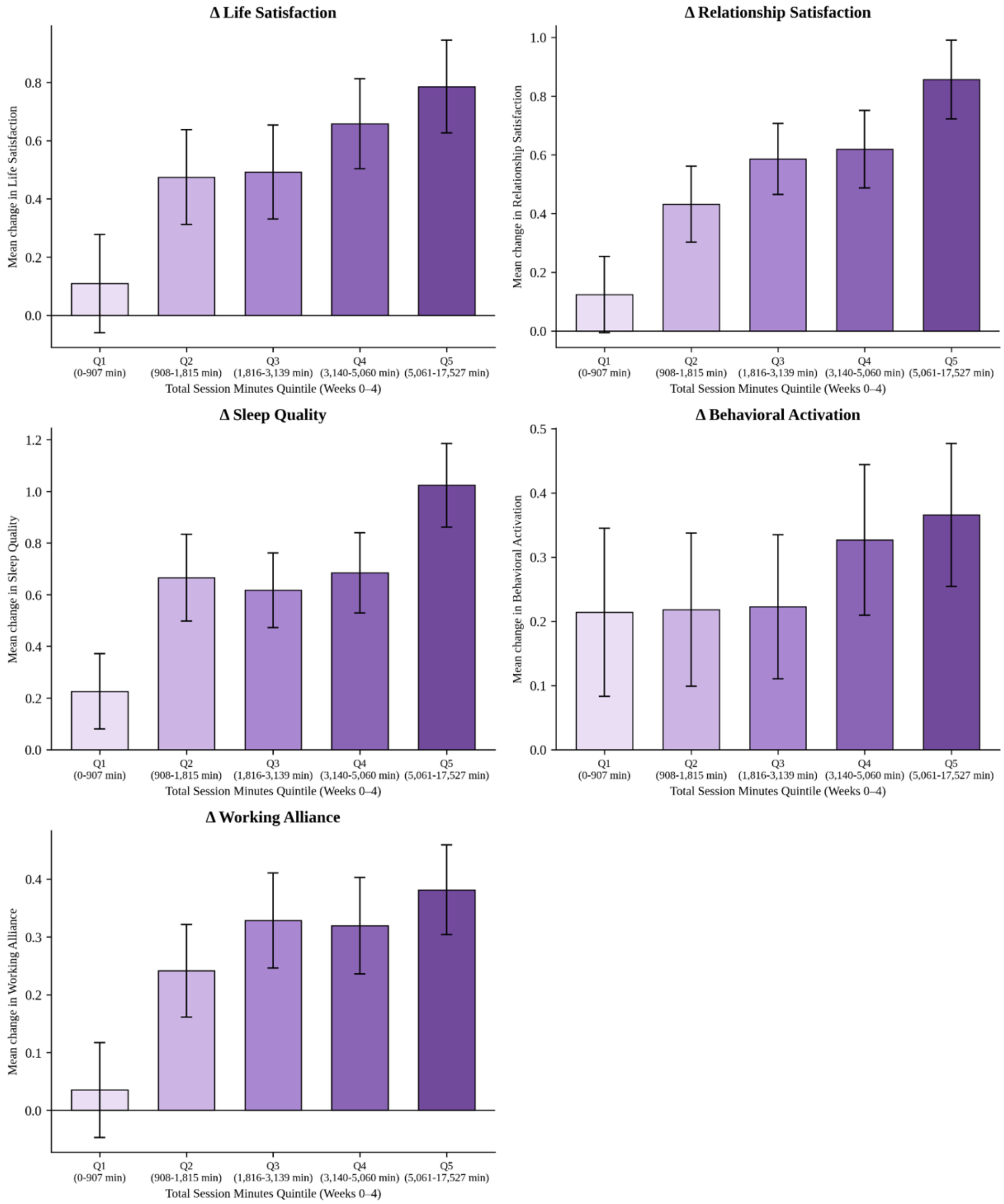

**Supplementary Figure 3 Caption:** *n* = 1,283. Bars represent mean change (Week 4 score minus baseline score) within each total session minutes quintile; error bars represent ±1 SE. Quintiles were derived from the distribution of total sessions minutes during the 28-day study period (Q1 = 0-907 minutes, Q2 = 908-1,815 minutes, Q3 = 1,816-3,139 minutes, Q4 = 3,140-5,060 minutes, Q5 = 5,061-17,527 minutes).